\documentclass[twocolumn,secnumarabic,amssymb, amsmath, nofootinbib,tightenlines,
nobibnotes, aps, prl,epsfig]{revtex4}
\usepackage{graphicx}% Include figure files
\usepackage{dcolumn}% Align table columns on decimal point
\usepackage{bm}% bold math
\begin{document}
\preprint{APS/123-QED}
\title{Top reduced cross section behavior at the LHeC kinematic range}% Force line breaks with \\

\author{G.R.Boroun}%
 \email{grboroun@gmail.com; boroun@razi.ac.ir }
\affiliation{ Physics Department, Razi University, Kermanshah
67149, Iran}% \textbackslash\textbackslash
\date{\today}% It is always \today, today,
             %  but any date may be explicitly specified
\begin{abstract}
%%%%%%%%%%%%%%%%%%%%%%%%%%%%%%%%%%%%%%%%%%%%%%%%%%%%%%%
The linear and non-linear behavior of the top reduced cross
section in the LHeC region is considered  concerning the
boson-gluon fusion (BGF) model at
 the deep inelastic scattering  leptoproduction. We will show that
 the nonlinear behavior ,at this region, for the $t\overline{t}$
 production is very small. Also the behavior of the top reduced cross section and  the ratio
 $R^{t}$  in this processes is considered.\\

%%%%%%%%%%%%%%%%%%%%%%%%%%%%%%%%%%%%%%%%%%%%%%%%%%%%%%%
\end{abstract}
 \pacs{***}%PACS, the Physics and Astronomy
                              %Classification Scheme.
 %Use showkeys class option if keyword
                              %display desired
\maketitle
%**********************************************************
%%%%%%%%%%%%%%%%%%%%%%%%%%%%%%%%%%%%%%%%%%%%%%%%%%%%%%%%%%%%%%%%%%%%%%%%%%%%%%%%%%%%%%%%%%%%%%
\subsection{1.Introduction}

Recently there have been important researches on the
$t\overline{t}$ production at the the Large Hadron Electron
Collider (LHeC) project at CERN [1-3]. The LHeC shows an increase
in the kinematic range of the deep inelastic scattering (DIS) and
also an increase in the luminosity since this value is
approximately $\simeq 100~fb^{-1}$. As to exceed, the integrated
HERA luminosity is realized by two orders of magnitude. This
continues the path of deep inelastic scattering  is the best tool
to probe proton structure into unknown areas of physics and DIS
kinematics. The DIS kinematics are $2 < Q^{2}< 100,000~GeV^{2}$
and $0.000002 < x < 0.8$ with a center-of-mass energy of about
$\sqrt{s_{ep}}>1~TeV$ in accordance with the B scenario, as the
kinematic range of the LHeC is illustrated in Fig.1. Clearly this
 increase in the precision of parton
distribution functions (PDF$^{,}$s) and in the low $x$ kinematic
region is realized by this project. As the strong rise of the
reduced cross section is expected to appear due to the non-linear
gluon-gluon interaction
effects in the so-called saturation region.\\
In the neutral current deep inelastic scattering (NC DIS), the
inclusive top electroproduction is dominantly by the boson-gluon
fusion at $x{\leq}0.1$ (Fig.2).  The top flavor production has not
been explored in NC DIS yet, because the cross section is too
small ( $\simeq 0.023 pb$). In the Standard Model, a precise
measurement of $t\overline{t}$ production in NC DIS  is sensitive
to the gluon distribution behavior [2], although the increase of
the gluon density towards low $x$ values must be tamed by the
unitarity reasons (
especially at "hot" spots in the proton).\\
In the framework of DGLAP dynamics [4-5],  considering the
heavy-flavor physics allowed a great advancement in understanding
of the heavy quark production dynamics.  New measurements of charm
and beauty production at LHeC provide high precision pQCD tests as
the LHeC is the ideal machine for a further extension on top
production. As to the measurements of LHeC, the events with charm
and beauty quarks increase of nearly  $\sim \%36$ and $\sim \%9$
fraction of the total respectively, when compared with HERA data
available. Due to the large gluon density at low $x$, the
$t\overline{t}$ pair production dominantly via the direct BGF
(Fig.2). In the high $Q^{2}$ limit, considering the top flavor
physics is in the framework of  fixed-flavor-number scheme (FFNS)
[6-7] with $n_{f}=5$. In general the measurement of the top
structure function $F_{2}^{t\overline{t}}$ is the highest interest
for theoretical analysis top quark in the final state. At
sufficiently high $Q^{2}>> m^{2}_{t}$ the top structure function
can be directly related to effective densities of top quark in the
proton.\\
The main purpose of this paper is the top reduced cross section
behavior in the LHeC regions. Top reduced cross section will be
presented in Sec.2. The nonlinear behavior is derived in Sec.3.
Sec.4 contains the conclusions and discussions.\\
\subsection{2.Top reduced  cross section}

We discuss the heavy flavor structure function
$F_{2}(x,Q^{2},m_{H}^{2})$, that is sensitive to the shape of the
gluon density at low $x$. The reaction under study is $e(p_{e}) +
P(p) {\rightarrow} e(p'_{ e}) + H(p_{1})\overline{H}(p_{1}) + X$,
where $H\overline{H}$ are heavy quark-antiquark pairs, such as
$c\overline{c},b\overline{b}$ or $t\overline{t}$, with momentum
$p_{1}$ $({p_{1}}^{2} = m_{H}^{2})$ and $X$ is any hadronic state
allowed. These pair productions are dependent to the momentum
carried by gluon distribution at low $x$. In the following, we are
interesting to determine the reduced cross section of the top pair
production
 in the LHeC region (Fig.1), although this cross section  in NC DIS is small [3].
 The top reduced cross section is given by the
 following form
\begin{eqnarray}
\widetilde{\sigma}^{t}(x,Q^{2},m_{t}^{2})=F_{2}^{t}(x,Q^{2},m_{t}^{2})[1-\frac{y^2}{1+(1-y)^{2}}R^{t}],
\end{eqnarray}
where $R^{t}=\frac{F_{L}^{t}}{F_{2}^{t}}$. The inclusive top
structure functions ($F_{2}^{t}$ and $F_{L}^{t}$) at the leading
order (LO) up to the next-to-leading order (NLO) analysis are
given in Ref. [8] (where only the gluon distributions are
considerable). The top structure functions are given by the
following forms
\begin{eqnarray}
F_{k}^{t}(x,Q^{2},m_{t}^{2}){\simeq}
\frac{Q^{2}\alpha_{s}}{4\pi^{2}m_{t}^{2}}\int^{z_{max}} _{x}
\frac{dz}{z}[ e_{t}^{2} g( \frac{x}{z}, \mu_{t}^{2})C^{(0)t}_{
k,g}\nonumber\\
 +
\frac{Q^{2}\alpha_{s}^{2}}{\pi m_{t}^{2}}\int^{z_{max}} _{x}
\frac{dz}{z}[ e_{t}^{2}g( \frac{x}{z}, \mu_{t}^{2})(C^{(1)t}_{
k,g}+\overline{C}^{(1)t}_{ k,g}\ln\frac{\mu_{t}^{2}}{m_{t}^{2}}),
\end{eqnarray}
where $k=2,L$ and the upper boundary on the integration is given
by $z_{max}=\frac{Q^{2}}{Q^{2}+4m^{2}_{t}}$, with $m_{t}=175~
GeV$. One can consider the boundary condition according to the
LHeC region, such that  $z_{max} {\rightarrow}~ 0.9$ when
$Q^{2}>4m^{2}_{t}$, $z_{max} {\rightarrow}~ 0.5$ when
$Q^{2}=4m^{2}_{t}$ and $z_{max} {\rightarrow}~ 0$ when
$Q^{2}<4m^{2}_{t}$. Here $g(x,\mu_{t}^{2})$ is the gluon density
and the scale $\mu_{t}(=\sqrt{\frac{Q^{2}}{2}+4m^{2}_{t}})$ is the
mass factorization and the renormalization scale. The
$\overline{MS}$ gluonic coefficient functions, $\{C^{(i)t}_{
k,g}(\eta,\xi), \overline{C}^{(i)t}_{ k,g}(\eta,\xi), i=0,1 \}$,
originate from those gluonic subprocesses where the virtual photon
is coupled to the top pair quark via the boson-gluon fusion and
they are dependent to the scaling variables
$\eta{\rightarrow}\frac{s}{4m_{t}^{2}}$ and
$\xi=\frac{Q^{2}}{m_{t}^{2}}$.\\
The top reduced cross section has been defined in Ref.[2]. Now,
let us introduce the compact formula for the ratio $R^{t}$ such
that
\begin{eqnarray}
R^{t}=\frac{C_{L,g}^{(i)t}(x,\xi){\otimes}g(x,\mu_{t}^{2})}{C_{2,g}^{(i)t}(x,\xi){\otimes}g(x,\mu_{t}^{2})},
\end{eqnarray}
here symbol $\otimes$ is correspondent to the convolution over the
variable $x$ as: $a \otimes b[x] = \int ^{1} _{x} \frac{dz}{z}
a(z)b(\frac{x}{z})$. The ratio
 $R^{t}$, in top-quark leptoproduction, is a probe of the top
 content of the proton which greatly simplifies the extraction of
 $F_{2}^{t}$
 from measurements of the top reduced cross sections at the LHeC. Fig.3 shows the quantities
$R^{c}, R^{b}$ and $R^{t}$ as a function of $Q^{2}$ for
$x{\leq}0.01$. The ratios $R^{c}$ and $R^{b}$ have approximately
the same behavior and they are difference from the ratio $R^{t}$
as these quantities are dependent to the gluon momentum, such that
$x_{g}^{t}>x_{g}^{b}>x^{c}_{g}$ (in accordance with the heavy
quark mass). These ratio$^{,}$s are approximately  independent of
$x$ at  low $x$ values. In Table 1, we see that $R^{max}\simeq
0.21$ for heavy quarks in a wide region of $Q^{2}$. We observe
that the $R^{max}$ predictions occurred in the region
$Q^{2}>4m^{2}_{H}$, as the ratio$^{,}$s $\frac{Q^{2}}{4m^{2}_{H}}$
are $11.11, 8.98$ and $8722.46$ for production charm, beauty and
top quarks respectively. One can consider the coefficient of
$R^{t}$ on a wide scale of the inelasticity $y$, such that
$\frac{y^2}{1+(1-y)^{2}}{\rightarrow}1$ at the LHeC region (Fig.1)
and according to the center of mass energy. Thus the top reduced
cross section can be rearranged by
\begin{eqnarray}
\widetilde{\sigma}^{t}(x,Q^{2},m_{t}^{2})=F_{2}^{t}(x,Q^{2},m_{t}^{2})[1-R^{t}].
\end{eqnarray}
At $Q^{2}\geq m^{2}_{t}$, the top structure function can be
evaluated from the top reduced cross section according to Eq.4 as
$F_{2}^{t}(x,Q^{2},m_{t}^{2})=
\widetilde{\sigma}^{t}(x,Q^{2},m_{t}^{2})/[1-R^{t}]$ which implies
that the $R^{t}$ values are given by Eq.3. The gluon distribution
to the top structure function is taken from MMHT14[10], NNPDF3[11]
and CT14[12] at large scale $\mu^{2}_{t}$. In what follows we need
an analytical form for the gluon distribution. So we shall use the
BDHM analysis [13] which extrapolated to the very low $x$ values
and high $Q^{2}$ values. The results ($F_{2}^{t}$ and
$\widetilde{\sigma}^{t}$) of our analysis at $Q^{2}\geq
m^{2}_{t}$, on the top electroproduction, are collected in Fig.4.
One can see that contribution of the top longitudinal structure
function $F_{L}^{t}$ to the top reduced
cross section will be measurable in the LHeC kinematic range.\\
For $Q^{2}<m^{2}_{t}$, the ratio $R^{t}$ is very small. So with a
good accuracy we will have
\begin{eqnarray}
\widetilde{\sigma}^{t}(x,Q^{2},m_{t}^{2})\simeq
F_{2}^{t}(x,Q^{2},m_{t}^{2}).
\end{eqnarray}
At low $x$, where only the gluon contributions are considerable,
the top contribution $F_{2}^{t}$ to the proton structure function
is given by this form
\begin{eqnarray}
F_{2}^{t}(x,Q^{2},m^{2}_{t})&=&2e_{t}^{2}\frac{\alpha_{s}(\mu^{2}_{t})}{2\pi}\int_{1-\frac{1}{a}}^{1-x}dzC_{2,g}^{(i)t}
(1-z,\frac{Q^{2}}{\mu_{t}^{2}})\nonumber\\
&& {\times}G(\frac{x}{1-z},\mu^{2}_{t}),
\end{eqnarray}
where $a=1+4\frac{m_{t}^{2}}{Q^{2}}$, $G=xg$ is the gluon
distribution function and $C_{2,g}^{(i)t}$$^{,}s$  are the top
coefficient functions at LO and NLO analysis. In what follows it
is convenient to use directly the gluon distribution behavior
according to the Eq.6 as the top reduced cross section (Eq.5)
modified by
\begin{eqnarray}
\widetilde{\sigma}^{t}(x,Q^{2},m_{t}^{2}){\propto}
C_{2,g}^{(i)t}(a_{s},x,\frac{Q^{2}}{\mu_{t}^{2}})
{\otimes}G(x,\mu_{t}^{2}).
\end{eqnarray}
This equation is valid when we will apply the high gluon density
 in accordance with the LHeC region at low $x$ values.  It provides a transition
from the dilute side by decreasing $x$ or by increasing the number
of gluons in the proton. We will study this transition from the
linear QCD evolution equations
into the non-linear equations in the next section.\\

%%%%%%%%%%%%%%%%%%%%%%%%%%%%%%%%%%%%%%%%%%%%%%%%%%%%%%%
\subsection{3.Nonlinearity}
 Within the standard framework of leading-twist linear
QCD evolution equations (DGLAP), the gluon densities  are
predicted to rise at low $x$ which is perceived in DIS experiments
at HERA. At the LHeC  low $x$ domains we have expected that the
growth of the top reduced cross section, with respect to the gluon
densities, saturates by unitarity bound (caused by Froissart and
Martin bounds [14]). This increase at low $x$ should eventually be
tames by the nonlinear effects of gluon density and it occurs when
chances of two gluons recombining into one and ,therefore, gluon
recombination will be  important concerning the gluon splitting.
This nonlinear evolution leads to the phenomenon of saturation of
gluonic density in the nucleon which can be exploited to the DIS
of the LHeC  at the high centre-of-mass energy(i.e.
extending the kinematic range to lower $x$).\\
This recombination leads to the modification of the linear DGLAP
evolution equation by a term which is nonlinear in gluon density.
The first equation of this type reporting the fusion of two gluon
ladders into on. This picture allows us to write the Gribov- Levin
-Ryskin- Mueller- Qiu (GLR-MQ) [15-16] equation for the gluon
distribution at low $x$. In double leading logarithmic
approximation (DLLA, $\ln{1/x}\ln{Q^{2}}>>1$), GLR-MQ  have been
determined a new nonlinear evolution equation for gluon density
because the negative sign in front of the nonlinear term
responsible to the gluon recombination, as:
\begin{eqnarray}
Q^{2}\frac{\partial^{2}G(x,Q^{2})}{{\partial}\ln{\frac{1}{x}}{\partial}Q^{2}}&=&\frac{\alpha_{s}N_{c}}{\pi}G(x,Q^{2})\nonumber\\
&&-\frac{4\alpha_{s}^{2}N_{c}}{3C_{F}R^{2}}\frac{1}{Q^{2}}[G(x,Q^{2})]^{2},
\end{eqnarray}
in which the parameter $R$ control the strength of the
nonlinearity and it is  the correlation radius between two
interaction gluons when
 that gluons are concentrated in
hot- spot [17] point $(R\simeq2\hspace{0.1cm} GeV^{-1})$ within
the proton.\\
Thus the $Q^{2}$ evaluation of the gluon distribution function
(Eq.8) with nonlinear effects, can be rewritten as
\begin{eqnarray}
\frac{{\partial}G(x,Q^{2})}{{\partial}{\ln}Q^{2}}&=&\frac{{\partial}G(x,Q^{2})}{{\partial}{\ln}Q^{2}}|_{DGLAP}\nonumber\\
&&-\frac{81\alpha^{2}_{s}}{16R^{2}Q^{2}}\int^{1}_{\chi}\frac{dy}{y}[G(y,Q^{2})]^{2},
\end{eqnarray}
where $\chi=\frac{x}{x_{0}}$ and  $x_{0}=0.01$ is the boundary
condition between the shadowing and unshadowing region that the
gluon distribution joints smoothly onto the unshadowed region
($x>x_{0}$). The first term in the r.h.s of Eq.9 is the standard
linear DGLAP evolution that can be expressed at low $x$ as
\begin{eqnarray}
\frac{{\partial}G(x,Q^{2})}{{\partial}{\ln}Q^{2}}|_{DGLAP}\simeq
\frac{\alpha_{s}}{2\pi}[P_{gg}(x,\alpha_{s}){\otimes}G(x,Q^{2})],
\end{eqnarray}
here the splitting function $P_{gg}$ can be ordered in running
coupling constant $\alpha_{s}(Q^{2})$. The second term in the
r.h.s of Eq.9 is the nonlinear  shadowing term. This shadowing
effects, arises from the two gluon ladders recombine into a single
one. Therefore the gluon distribution function modified by the
following form [18-19]
\begin{eqnarray}
\frac{{\partial}G(x,Q^{2})}{{\partial}{\ln}Q^{2}}&=&\frac{\alpha_{s}}{2\pi}P_{gg}(x,\alpha_{s}){\otimes}G(x,Q^{2})\nonumber\\
&&-\frac{81\alpha_{s}^{2}}{16R^{2}Q^{2}}\int^{1}_{\chi}\frac{dy}{y}[G(y,Q^{2})]^{2}.
\end{eqnarray}
The solution of Eq.11 is straightforward and given by
\begin{eqnarray}
G(x,Q^{2})=\frac{M(x,Q^{2})}{C-N(x,Q^{2})\Gamma[-1+M(x),Q^{2}]},
\end{eqnarray}
in which the functions of $M$ and $N$  are dependent on the
coefficients of $G(x,Q^{2})$ and  $G^{2}(x,Q^{2})$ in the r.h.s of
Eq.11 respectively, also $\Gamma$ is the incomplete gamma function
and $C$ is dependent to the initial condition of the gluon
distribution at $Q^{2}=Q_{0}^{2}$. The shadowing correction to the
gluon distribution at the initial scale defined by
$G(x,Q_{0}^{2})$. As, the low $x$ behavior of the shadowing
corrections to the gluon distribution at $Q_{0}^{2}$ is assumed to
be [20]
\begin{eqnarray}
G(x,Q_{0}^{2})=G^{u}(x,Q_{0}^{2})[1+\theta(x_{0}-x)[G^{u}(x,Q_{0}^{2})\nonumber\\
-{G}^{u}(x_{0},Q_{0}^{2})]/G^{sat}(x,Q_{0}^{2})]^{-1},
\end{eqnarray}
where
$G^{sat}(x,Q^{2})=\frac{16R^{2}Q^{2}}{27\pi\alpha_{s}(Q^{2})}$ is
the value of the gluon which would saturate the unitarity limit in
the leading shadowing approximation and $G^{u}(x,Q_{0}^{2})$ that
is the input unshadowing gluon distribution taken from the QCD
parametrisation.\\
Now, we have  the shadowing correction to the initial gluon
distribution, then inserting Eq.13 in Eq.12, therefore the C
constant in Eq.12 is dependence upon the initial condition as it
follows:
\begin{widetext}
\begin{eqnarray}
G(x,Q^{2})=~~~~~~~~~~~~~~~~~~~~~~~~~~~~~~~~~~~~~~~~~~~~~~~~~~~~~~~~~~~~~~~~~~\nonumber\\
 \frac{M(x,Q^{2})G^{u}(x,Q_{0}^{2})}
{M(x,Q_{0}^{2})[1+\frac{\theta(x_{0}-x)[G^{u}(x,Q_{0}^{2})-{G}^{u}(x_{0},Q_{0}^{2})]}{G^{sat}(x,Q_{0}^{2})}]+\{N(x,Q_{0}^{2})\Gamma[-1+M(x),Q_{0}^{2}]
-N(x,Q^{2})\Gamma[-1+M(x),Q^{2}]\}G^{u}(x,Q_{0}^{2})}.\nonumber\\
\end{eqnarray}
\end{widetext}
 Therefore the shadowing correction to the top reduced cross section is
corespondent to the nonlinear gluon distribution function at scale
$\mu_{t}^{2}$ by the following form:
\begin{widetext}
\begin{eqnarray}
\widetilde{\sigma}^{t}(x,Q^{2},m_{t}^{2})\simeq~
2e_{t}^{2}\frac{\alpha_{s}(\mu^{2}_{t})}{2\pi}C_{2,g}^{(i)t}(a_{s},x,\frac{Q^{2}}{\mu_{t}^{2}})
{\otimes}~~~~~~~~~~~~~~~~~~~~~~~~~~~~~~~~~~~~~~~~~~~~~~~~~~~~~~~~~~~~~~~~~~~~~~~~~~~~~~~~~~~~~~~~~~~~\nonumber\\
 \frac{M(x,\mu_{t}^{2})G^{u}(x,\mu_{0t}^{2})}
{M(x,\mu_{0t}^{2})[1+\frac{\theta(x_{0}-x)[G^{u}(x,\mu_{0t}^{2})-{G}^{u}(x_{0},\mu_{0t}^{2})]}{G^{sat}(x,\mu_{0t}^{2})}]+\{N(x,\mu_{0t}^{2})
\Gamma[-1+M(x),\mu_{0t}^{2}]
-N(x,\mu_{t}^{2})\Gamma[-1+M(x),\mu_{t}^{2}]\}G^{u}(x,\mu_{0t}^{2})}.\nonumber\\
\end{eqnarray}
\end{widetext}
This equation shows that the top reduced cross section behavior is
tamed due to the saturation effects at low $x$. At the LHeC region
for $Q^{2}<1000~GeV^{2}$ we observed that $\mu_{t}^{2} \simeq
4m_{t}^{2}$, therefore the coefficient of the nonlinear term
(second term in the r.h.s of Eq.9) is very very small and the
shadowing correction to the top reduced cross section is not
worthwhile. We believe that this behavior for the nonlinear term
at low $Q^{2}$ values is correct, because within the color dipole
approach [21] the $t\overline{t}$ color dipole size is very small
than the charm and bottom size ($0.037 \leq r_{c} \leq 0.130 fm $
and $0.014 \leq r_{b} \leq 0.043 fm $) as
\begin{eqnarray}
\frac{4}{Q^{2}+4m^{2}_{t}} \leq r_{t}^{2} \leq \frac{1}{m^{2}_{t}}
{\Rightarrow} ~r_{t}\simeq ~0.0013~ fm.
\end{eqnarray}
We observe that the minimum and maximum distances, related to the
 $t\overline{t}$ interaction in the color dipole model, are the same one value
$0.0013~fm$. The color dipole size shows the sticking of the
$t\overline{t}$ pair production, therefore in this case the
nonlinear effects can be neglected for the LHeC region. So we can
conclude that the shadowing corrections to the top reduced cross
sections  are very small in accordance with the LHeC region. On
the other hand, in the region of low- $x$ and  $Q^{2}$ values, the
process of gluon recombination does not play an important role on
the $t\overline{t}$ production and therefore the nonlinear
corrections to the top reduced cross section is not essential at
this region. In Fig.5 we show our results  for $Q^{2}=100
~GeV^{2}$, as the top
reduced cross sections  are very small for measurements.\\

\subsection{4.Conclusion }
We have obtained the ratio $R^{t}={F_{2}^{t}}/{F_{L}^{t}}$ at the
LHeC regions for the top pair production.  We demonstrated the
relation
 between the top reduced cross section and the top structure
 function by this ratio ($R^{t}$) at $Q^{2} \geq m_{t}^{2}$ and $Q^{2} <
 m_{t}^{2}$ regions. The linear and nonlinear behaviors considered and shows that the nonlinear behavior has a very
small content for prob at very low $x$ in which the gluon
saturation is dominance.\\

%%%%%%%%%%%%%%%%%%%%%%%%%%%%%%%%%%%%%%%%%%%%%%%%%%%%%%%
\subsection{Acknowledgments}
 Author is grateful to Prof.N.Armesto for suggestion, reading the manuscript and useful comments.\\

%%%%%%%%%%%%%%%%%%%%%%%%%%%%%%%%%%%%%%%%%%%%%%%%%%%%%%%%%%%%%
\section{References}
1. LHeC workshops 2015 (http://cern.ch/lhec).\\
2.G.R.Boroun, Phys.Lett.B{\bf744} (2015)142; Phys.Lett.B{\bf741} (2015)197. \\
3. J.L.Abelleira Fernandez, et.al., [LHeC Collab.],
J.Phys.G\textbf{39} (2012)075001.\\
4. V.N. Gribov and L.N. Lipatov, Sov. J. Nucl. Phys.{\bf18}
(1972) 438.\\
5. L.N. Lipatov, Sov. J. Nucl. Phys.{\bf20} (1975) 93; G.
Altarelli and G. Parisi, Nucl. Phys. B{\bf126} (1977) 298; Yu.L.
Dokshitzer, Sov. Phys. JETP {\bf46} (1977) 641.\\
6. M.A.G.Aivazis, et.al., Phys.Rev.D{\bf50} (1994) 3102.\\
7. J.C.Collins, Phys.Rev.D{\bf58} (1998) 094002.\\
8.E. Laenen, S. Riemersma, J. Smith and W.L. van Neerven, Nucl.
Phys. B{\bf392} (1993) 162; S. Riemersma, J. Smith and W.L. van
Neerven, Phys.Lett.B{\bf347} (1995) 143.\\
9. A.~Y.~Illarionov,B.~A.~Kniehl and A.~V.~Kotikov, Phys.Lett. B {\bf 663}  (2008)66.\\
10. L.A Harland-Lang, et.al., Eur.Phys.J.C{\bf75} (2014) 435.\\
11.  Richard D. Ball, et.al., NNPDF Collaboration, JHEP{\bf04} (2015) 040.\\
12. S.Dulat, et.al., arXiv:1506.07443 [hep-ph] (2016).\\
13. Martin M.Block, et.al., Phys. Rev. D{\bf 88} (2013) 014006;
Phys. Rev. D{\bf 88} (2013) 013003.\\
 14. M. Froissart, Phys. Rev. {\bf123} (1961) 1053; A. Martin,
Phys. Rev. {\bf129} (1963) 1432.\\
15. L.V.Gribov, E.M.Levin and M.G.Ryskin, Phys.Rep.\textbf{100}
 (1983)1.\\
16. A.H.Mueller and J.Qiu, Nucl.Phys.B\textbf{268} (1986)427.\\
17. E.M.Levin and M.G.Ryskin, Phys.Rep.\textbf{189} (1990)267.\\
18. G.R.Boroun, JETP{\bf106}(2008)701; G.R.Boroun and B.Rezaei,
Eur.Phys.J.C{\bf73} (2013)2412.\\
 19. G.R.Boroun, Eur.Phys.J.A{\bf42} (2009)251; G.R. Boroun and S. Zarrin,
Eur.Phys.J.Plus{\bf128} (2013)119.\\
20. A.D. Martin, W.J. Stirling, R.G. Roberts, R.S. Thorne, Phys.
Rev. D {\bf47} (1993) 867; J. Kwiecinski, A.D. Martin, P.J.
Sutton, Phys. Rev. D{\bf 44} (1991) 2640; A.J. Askew, J.
Kwiecinski, A.D. Martin, P.J. Sutton, Phys. Rev. D{\bf 47} (1993)
3775.\\
21.R.Fiore, N.N.Nikolaev and V.R.Zoller, JETP Lett.{\bf90}
(2009)319.\\
%%%%%%%%%%%%%%%%%%%%%%%%%%%%%%%%%%%%%%%%%%%%%%%%%%%%%%%%%%%%%%%%

%%%%%%%%%%%%%%%%%%%%%%%%%%%%%%%%%%%%%%%%%%%%%%%%%%%%%%%%%%
\begin{table}
\centering \caption{Maximum values of $R^{H}$ extracted from Eq.3.
}\label{table:table1}
\begin{minipage}{\linewidth}
\renewcommand{\thefootnote}{\thempfootnote}
\centering
\begin{tabular}{|l|c|c|c|l|} \hline\noalign{\smallskip} $Q^{2}(GeV^{2})$ & $R_{max}^{c}$ & $R_{max}^{b}$ & $R_{max}^{t}$ \\
\hline\noalign{\smallskip}
100 & 0.2106 & --- & ---  \\
750 & ---  & 0.2086 & --- \\
860000 & ---   & --- & 0.2093 \\

 \hline\noalign{\smallskip}
\end{tabular}
\end{minipage}
\end{table}
%%%%%%%%%%%%%%%%%%%%%%%%%%%%%%%%%%%%%%%%%%%%%%%%%%%%%%%%%%

\begin{figure}
\includegraphics[width=0.30\textwidth]{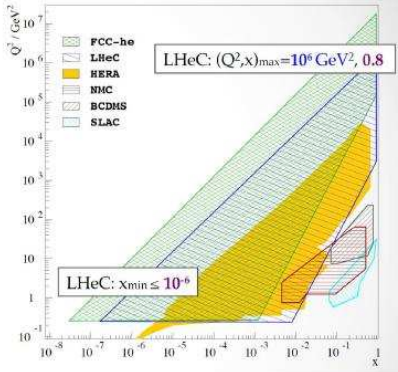}
\caption{Kinematic plane ($Q^{2},x$) at the LHeC region.}
\end{figure}

\begin{figure}
\includegraphics[width=0.25\textwidth]{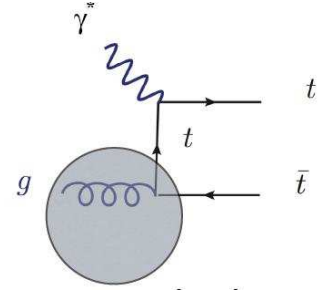}
\caption{Leading order Boson Gluon Fusion (BGF) diagram for
$t\overline{t}$ production in ep-collisions.}
\end{figure}

\begin{figure}
\includegraphics[width=0.5\textwidth]{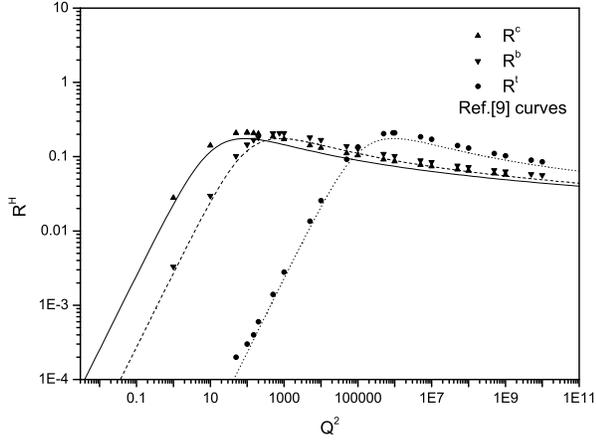}
\caption{$R^{H}$ $^{,}$s evaluated as functions of $Q^{2}$ at NLO
analysis with $<\mu_{H}^{2}>=4m^{2}_{H}+\frac{Q^{2}}{2}$, compared
with curves estimated in Ref.[9].}
\end{figure}

\begin{figure}
\includegraphics[width=0.5\textwidth]{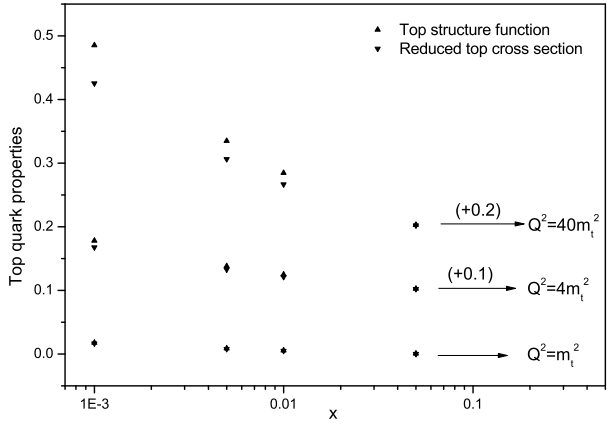}
\caption{Top structure function $F_{2}^{t}$ and top reduced  cross
section $\widetilde{\sigma}^{t}$ determined from Eq.4 at
$Q^{2}=m_{t}^{2}, 4m_{t}^{2}$ and $40m_{t}^{2}~GeV^{2}$ according
to the LHeC region.}
\end{figure}

\begin{figure}
\includegraphics[width=0.5\textwidth]{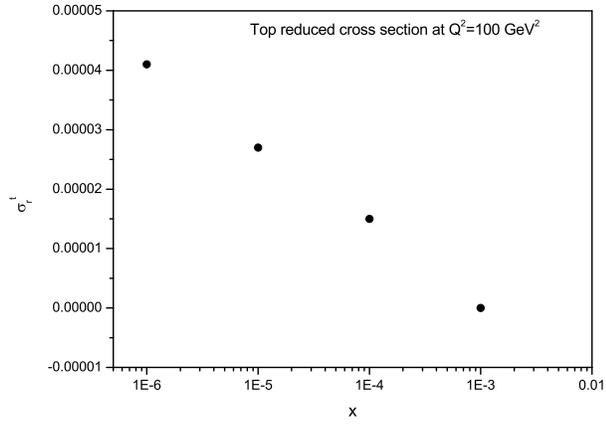}
\caption{Top reduced  cross section $\widetilde{\sigma}^{t}$
determined from Eq.15 at $Q^{2}=100~GeV^{2}$ according to the LHeC
region.}
\end{figure}
%%%%%%%%%%%%%%%%%%%%%%%%%%%%%%%%%%%%%%%%%%%%%%%%%%%%%%%%%%

%%%%%%%%%%%%%%%%%%%%%%%%%%%%%%%%%%%%%%%%%%%%%%%%%%%%%%%%%%

\end{document}